\documentclass[pre,aps,floats,twocolumn,showpacs,superscriptaddress,floatfix]{revtex4}
\usepackage{graphicx,amssymb,amsmath,ifthen}
\usepackage[active]{srcltx}
\begin{document}
%
\title{Quantum integrability and nonintegrability in the spin-boson model}
%
\author{Vyacheslav V. Stepanov} 
\affiliation{Department of Physics, University of Rhode Island, Kingston RI
  02881, USA} 
\author{Gerhard M{\"u}ller} 
\affiliation{Department of Physics, University of Rhode Island, Kingston RI
  02881, USA}
\author{Joachim Stolze} 
\affiliation{Institut f{\"u}r Physik, Technische Universit{\"a}t Dortmund,
 44221 Dortmund, Germany}
\date{\today}
\begin{abstract}
  We study the spectral properties of a spin-boson Hamiltonian that depends on
  two continuous parameters $0\leq\Lambda<\infty$ (interaction strength) and
  $0\leq\alpha\leq\pi/2$ (integrability switch). In the classical limit this
  system has two distinct integrable regimes, $\alpha=0$ and $\alpha=\pi/2$.
  For each integrable regime we can express the quantum Hamiltonian as a
  function of two action operators.  Their eigenvalues (multiples of $\hbar$)
  are the natural quantum numbers for the complete level spectrum. This
  functional dependence cannot be extended into the nonintegrable regime
  $(0<\alpha<\pi/2)$. Here level crossings are prohibited and the level
  spectrum is naturally described by a single (energy sorting) quantum number.
  In consequence, the tracking of individual eigenstates along closed paths
  through both regimes leads to conflicting assignments of quantum numbers.
  This effect is a useful and reliable indicator of quantum chaos -- a
  diagnostic tool that is independent of any level-statistical analysis.
\end{abstract}


\maketitle

%
\section{Introduction}\label{sec1}
%

Classical integrability of a system with two degrees of freedom guarantees that
the Hamiltonian can be expressed as a piecewise smooth function of two action
coordinates: $H(p_1,q_1;p_2,q_2) = H_C(J_1,J_2)$. No such functional relation
exists if the system is nonintegrable \cite{Tabo89, Gutz90, Reic92, Hilb00}.
Geometrically speaking for a parametric system with parameters subject to an
integrability condition, there exist complete foliations of invariant tori in
phase space for all parameter points in the integrable regime. Throughout the
nonintegrable regime the foliation is partially destroyed. Some tori are
replaced by chaotic trajectories, cantori, and unstable periodic trajectories.
The surviving tori in the nonintegrable regime are no longer dense anywhere in
phase space, but each one of them can still be characterized by two local
action coordinates $J_1,J_2$ via line integrals $\oint p_{i}dq_{i}$ along pairs
of topologically independent closed paths.

In the integrable regime, the natural label of an invariant torus is its set of
action coordinates $(J_1,J_2)$. Tracking an invariant torus along a path
through the integrable regime of parameter space means that we observe how the
torus with this specification changes its position and shape in phase space.
In the nonintegrable regime, where all intact tori are separated by chaotic
phase flow, an individual torus can no longer be characterized by fixed values
of $J_1,J_2$.  Tracking a surviving invariant torus along a path through the
nonintegrable regime of parameter space now means that we observe it in
isolation from other tori. The local action coordinates vary smoothly as the
torus changes its location and shape in phase space.

Now let us attempt to track one torus along a closed path that lies partly in
the integrable regime and partly in the nonintegrable regime, assuming that it
does survive the presence of chaos. Inside the integrable regime the identity
of the torus is determined by constant values of the action coordinates, while
outside that regime the action coordinates vary with the shape of the isolated
torus. The values of $J_1,J_2$ at the end of the closed path will, in general,
be different from the starting values, implying that the individuality of a
torus cannot be maintained. No loss of individuality is suffered by tori along
closed paths embedded in the integrable regime or for surviving tori along
closed paths in the nonintegrable regime \cite{Note1}.

There exists a quantum counterpart to this 'crisis of identity' as will be
demonstrated. It can be employed to discriminate between regimes of
integrability and nonintegrability on purely quantum mechanical grounds. Here
we show the workings of this diagnostic tool in the context of the
spin-boson model \cite{CCW+90, CCM+91, MSNL91, CCM95},
\begin{align}\label{hamil}
H &= \hbar\omega_Ba^{\dagger}a+\hbar\omega_SS_z \nonumber \\
  &+ \Lambda\cos\alpha\left(S_{+}a+S_{-}a^{\dagger}\right) \nonumber \\
  & + \Lambda\sin\alpha\left(S_{+}a^{\dagger }+S_{-}a\right),
\end{align}
one of the simplest nontrivial models describing nonrelativistically the
interaction between an atom and a radiation field \cite{Note2}. This model has
also been used to describe the interaction between electronic and vibrational
degrees of freedom in molecules and solids. The relation between classical and
quantum integrability of (\ref{hamil}) has been the object of previous
investigations \cite{GH84, MSNL91}.

The first two terms in (\ref{hamil}) describe one mode of the electromagnetic
field and a $(2\sigma+1)$-level atom, respectively. The coupling between the
two degrees of freedom has strength $\Lambda$ and depends on a continuous
parameter $\alpha$ that connects two regimes for which this model is integrable
in the classical limit. The classical integrability for $\alpha=0$ and
$\alpha=\pi/2$ is established by a second integral of the motion. The case
$\alpha=0$ is known as the rotating wave approximation in quantum optics. Early
studies in one or the other classical limit of the spin-boson model revealed
chaotic phase space flow turning regular in the rotating wave approximation
\cite{CCW+90, CCM+91, BZT76, MAG83}.

In the two-dimensional parameter space spanned by the (polar) coordinates
$(\Lambda,\alpha)$, the two integrable regimes are located on two perpendicular
straight lines that intersect each other at the point of zero coupling
strength.  Each quadrant of this parameter plane represents a nonintegrable
regime. Henceforth we consider the parameter range $0 \leq\Lambda<\infty$,
$0\leq\alpha\leq\pi/2$.

In preparation of our main theme we first discuss the classical integrability
condition of the spin-boson model (Sec.~\ref{sec22}) and then the
classification of its quantum energy levels (Sec.~\ref{sec2}) and certain
quantum invariants (Sec.~\ref{sec3}) by distinct sets of quantum numbers in the
integrable and nonintegrable regimes. This distinction has a deeper meaning,
which we will further discuss in Sec.~\ref{sec4}, and which we will employ in
Sec.~\ref{sec5} for the identification of the two regimes in purely quantum
mechanical terms.

%
\section{Integrability Condition}\label{sec22}
%

In taking the classical limit $\hbar\to 0$, $\sigma\to\infty$ of the spin-boson
model, we renormalize the coupling constant,
$\Lambda=(\hbar/2)^{3/2}\bar{\Lambda}$, substitute 
\begin{equation}
  \label{eq:1}
  a = \sqrt{\frac{M\omega_B}{2\hbar}}\,x +
\sqrt{\frac{1}{2\hbar M\omega_B}}\,\imath p
\end{equation}
for the boson operators and, via
$\hbar\sqrt{\sigma(\sigma+1)}=s$, convert the spin-$\sigma$ operator into a
classical 3-component vector of fixed length: 
\begin{equation}
  \label{eq:2}
  (S_x,S_y,S_z) =
s(\sin\vartheta\cos\varphi, \sin\vartheta\sin\varphi, \cos\vartheta).
\end{equation}
The spin-boson Hamiltonian (\ref{hamil}) thus turns into the energy function of
two linear one-degree-of-freedom systems -- a harmonic oscillator and a
classical spin in a constant magnetic field -- with a nonlinear coupling:
\begin{align}\label{hcl}
H &= \frac{p^2}{2M} + \frac{1}{2}M\omega_B^2x^2 + \omega_SS_z \nonumber \\
&+ \frac{1}{2}\bar{\Lambda}\cos\alpha\left[\sqrt{M\omega_B}xS_x 
- \frac{1}{\sqrt{M\omega_B}}pS_y\right] \nonumber \\
&+ \frac{1}{2}\bar{\Lambda}\sin\alpha\left[\sqrt{M\omega_B}xS_x 
+ \frac{1}{\sqrt{M\omega_B}}pS_y\right].
\end{align}
A set of canonical coordinates is $(p,x;s\cos\vartheta,\varphi)$. The equations
of motion for the physical variables $(x,p,S_x,S_y,S_z)$ inferred from
(\ref{hcl}) via $dx/dt= \partial H/\partial p$, $dp/dt= -\partial H/\partial x$,
and $d{\bf S}/dt= -{\bf S}\times \partial H/\partial{\bf S}$ read
\begin{subequations}\label{jc}
\begin{align}
\dot x &= \frac{p}{M} + \frac{\bar{\Lambda}}{2\sqrt{M\omega_B}}S_y
(\sin\alpha - \cos\alpha), \\
\dot p &= - M\omega^2_B x - \frac{\bar{\Lambda}\sqrt{M\omega_B}}{2}
S_x(\cos\alpha + \sin\alpha), \\
\dot{S}_x &= -\omega_S S_y - 
\frac{\bar{\Lambda} p}{2 \sqrt{M\omega_B}} S_z (\cos\alpha - \sin\alpha), \\
\dot{S}_y &= \omega_S S_x - 
\frac{\bar{\Lambda} x \sqrt{M\omega_B}}{2} S_z (\cos\alpha + \sin\alpha), \\
\dot{S}_z &= 
\frac{\bar{\Lambda} x \sqrt{M\omega_B}}{2} S_y (\cos\alpha + \sin\alpha) \nonumber \\
& \hspace*{1.5cm} 
+ \frac{\bar{\Lambda} p}{2\sqrt{M\omega_B}} S_x (\cos\alpha - \sin\alpha). 
\end{align}
\end{subequations}

The phase flow generated by these equations is, in general, chaotic. Chaos gives
way to a fully intact torus structure at $\alpha=0,\pi/2$. The integrability of these
cases is established by the fact that one or the other of the two functions,
\begin{subequations}\label{cl-inv}
\begin{align}
I &= \frac{p^2}{2M\omega_B} + \frac{1}{2}M\omega_Bx^2 + S_z, \\
K &= \frac{p^2}{2M\omega_B} + \frac{1}{2}M\omega_Bx^2 - S_z,
\end{align}
\end{subequations}
whose time evolution is determined by \cite{note1}
\begin{subequations}\label{IdotKdot}
\begin{align}
{\dot I} = \{H,I\} &= \bar{\Lambda}\sin\alpha\left(\frac{pS_x}{\sqrt{M\omega_B}} - 
         \sqrt{M\omega_B}xS_y\right), \\
{\dot K} = \{H,K\} &= \bar{\Lambda}\cos\alpha\left(\frac{pS_x}{\sqrt{M\omega_B}} + 
         \sqrt{M\omega_B}xS_y\right),
\end{align}
\end{subequations}
becomes a second integral of the motion. The case $\alpha=0$ is known as the
Jaynes-Cummings model \cite{JC63}. The impact of the classical integrability
conditions on the quantum system is the main theme of this study.

%
\section{Energy Levels}\label{sec2}
%

\begin{figure}[tb]
\vspace*{0.4cm}
\centerline{\includegraphics[width=8.0cm]{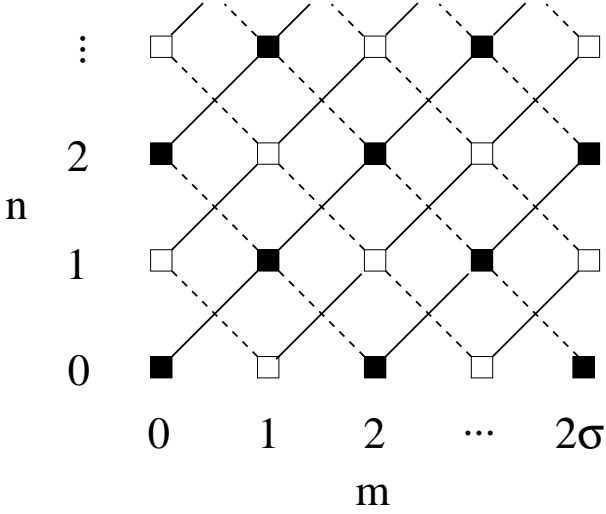}}
\vspace*{0.6cm}
\caption{Basis vectors $|m,n\rangle$ with positive parity (full squares) and negative
  parity (open squares) as coupled by matrix elements of operators
  $S_+a,S_-a^\dagger$ (solid lines) and $S_+a^\dagger,S_-a$ (dashed
  lines) of Hamiltonian (\ref{hamil})}
\label{fig2}
\end{figure}

For the analytic or numerical solution of the spin-boson model (\ref{hamil}),
it is convenient to use the product vectors of the noninteracting system,
$|m,n\rangle,~ m=0,1,2,\ldots,2\sigma,~ n=0,1,2,\ldots$, as a basis. The
relevant operators act on this basis as follows:
\begin{subequations}\label{opbas}
\begin{align}
(\sigma-S_z)|m,n\rangle &= m|m,n\rangle, \\
S_+|m,n\rangle &= \sqrt{m(2\sigma-m+1)}|m-1,n\rangle, \\ 
S_-|m,n\rangle &= \sqrt{(2\sigma-m)(m+1)}|m+1,n\rangle, \\ 
a^\dagger|m,n\rangle &= \sqrt{n+1}|m,n+1\rangle, \\
a|m,n\rangle &= \sqrt{n}|m,n-1\rangle.
\end{align}
\end{subequations}
The Hamiltonian matrix can thus be assembled from the diagonal elements
\begin{equation}\label{on-dia}
\langle m,n|S_z|m,n\rangle = \sigma-m,~ ~ \langle m,n|a^\dagger a|m,n\rangle =n,
\end{equation}
and from the off-diagonal elements
\begin{align}
\langle m,n|S_+a|m+1,n+1\rangle &= \sqrt{(2\sigma-m)(m+1)(n+1)},
\nonumber \\
\langle m,n|S_-a^\dagger|m-1,n-1\rangle &= \sqrt{(2\sigma+1-m)mn},
\nonumber \\
\langle m,n|S_+a^\dagger|m+1,n-1\rangle &= \sqrt{(2\sigma-m)(m+1)n}, 
\nonumber \\
\langle m,n|S_-a|m-1,n+1\rangle &= \sqrt{(2\sigma+1-m)m(n+1)}.
\nonumber
\end{align}
The structure of this matrix is illustrated in Fig.~\ref{fig2}. The solid lines
represent matrix elements generated by the first interaction term in
(\ref{hamil}), and the dashed lines represent matrix elements which arise in the
second interaction term.
The two-sublattice structure is a reflection of parity conservation. The parity
operator,
\begin{equation}\label{pari}
P = (-1)^{a^\dagger a + \sigma - S_z},
\end{equation}
commutes with $H$ for arbitrary $\Lambda,\alpha$. It divides the
eigenstates into two symmetry classes. States with $P=+1$ $(P=-1)$ involve
basis vectors with even $m+n$ (odd $m+n$) only. 

If $\alpha=0$ only the solid bonds are present and if $\alpha=\pi/2$ only the
dashed bonds. In either case the Hamiltonian matrix is reduced to invariant
blocks of size $2\sigma+1$. If $0<\alpha<\pi/2$ we must deal with an infinite
matrix. In this study we set $\omega_S=\omega_B\equiv\omega$ except where
indicated otherwise. In the following we analyze the level spectrum for various
cases at $\alpha=0,\pi/2$: systems with $\sigma=\frac{1}{2}$, $\sigma=1$ for
arbitrary $n$, and systems with arbitrary $\sigma$ asymptotically for large
$n$.

\subsection{Spin $\sigma=\frac{1}{2}$}\label{spinh}

The integrable spin-boson model with $\sigma=\frac{1}{2}$ involves only $2\times
2$ matrices. If $\alpha=0$, the eigenvectors happen to be independent of
the interaction strength,
\begin{subequations}\label{psi1}
\begin{align}
|\psi_{1,0}\rangle &= |1,0\rangle \\
|\psi_{1,n}\rangle &= \frac{1}{\sqrt{2}}\left\{|1,n\rangle 
                    + |0,n-1\rangle\right\},~n=1,2,\ldots, \\
|\psi_{0,n}\rangle  &= \frac{1}{\sqrt{2}}\left\{|1,n+1\rangle 
                    - |0,n\rangle\right\},~n=0,1,\ldots,  
\end{align}
\end{subequations}
and the energy eigenvalues (for $n=0,1,2,\ldots$) are
\begin{subequations}\label{emn1}
\begin{align}
E_{1,n} &= \hbar\omega(n-1/2) + \Lambda\sqrt{n}, \\
E_{0,n} &= \hbar\omega(n+1/2)- \Lambda\sqrt{n+1}.
\end{align}                
\end{subequations}
If $\alpha=\pi/2$ the eigenvectors do depend on $\Lambda$:
\begin{subequations}\label{psi2}
\begin{align}
|\psi_{1,0}\rangle &= |0,0\rangle \\
|\psi_{1,n}\rangle &= a_{0,n}|0,n\rangle 
                    + b_{0,n}|1,n-1\rangle,~n=1,2,\ldots, \\
|\psi_{0,n}\rangle  &= a_{1,n}|0,n+1\rangle
                     - b_{1,n+1}|1,n\rangle,~n=0,1,\ldots
\end{align}
\end{subequations}
with
\begin{align*}
a_{0,n} &= \frac{\sqrt{\lambda_n-1}}{\sqrt{2(\lambda_1-\sqrt{\lambda_n})}},\quad 
a_{1,n-1} = \frac{\sqrt{\lambda_n-1}}{\sqrt{2(\lambda_1+\sqrt{\lambda_n})}}, \\
b_{0,n} &= \frac{\sqrt{\lambda_n}-1}{\sqrt{2(\lambda_1-\sqrt{\lambda_n})}},\quad
b_{1,n-1} = \frac{\sqrt{\lambda_n}+1}{\sqrt{2(\lambda_1+\sqrt{\lambda_n})}},
\end{align*}
where $\lambda \doteq (\Lambda/\hbar\omega)^2$, $\lambda_n \doteq 1+n\lambda$.
The associated energy eigenvalues (for $n=0,1,2,\ldots$) are
\begin{subequations}\label{emn2}
\begin{align}
E_{1,n} &= \hbar\omega(n-1/2)
         + \hbar\omega\sqrt{\lambda_n}, \\
E_{0,n} &= \hbar\omega(n+1/2)
         - \hbar\omega\sqrt{\lambda_{n+1}}.    
\end{align}
\end{subequations}

\subsection{Spin $\sigma=1$}\label{spin1}

The case $\sigma=1$ at integrability involves the solution of cubic
equations. Here we list the ($\Lambda$-independent) eigenvectors and the
associated energy eigenvalues for $\alpha=0$. We have 
$|\psi_{1,0}\rangle = |0,0\rangle$, $|\psi_{1,1}\rangle = (|1,0\rangle +
|0,1\rangle)/\sqrt{2}$, $|\psi_{2,1}\rangle = (|1,0\rangle -
|0,1\rangle)/\sqrt{2}$, with energies 
$E_{1,0} = -\hbar\omega$, $E_{1,1} = \sqrt{2}\Lambda$, 
$E_{2,1} = -\sqrt{2}\Lambda$, respectively, and for $n\geq 2$ the
results are
\begin{subequations}\label{evpi2}
\begin{align}
\left| \psi_{1,n} \right> &= \sqrt{\frac{n-1}{4n-2}} \left| 0,n-2
 \right> +
\frac{1}{\sqrt{2}} \left| 1,n-1 \right> \nonumber \\ &+
\sqrt{\frac{n}{4n-2}} \left| 2,n \right>, \\
\left| \psi_{2,n} \right> &= \sqrt{\frac{n}{2n-1}} \left| 0,n-2 \right> +
\sqrt{\frac{n-1}{2n-1}} \left| 2,n \right>, \\
\left| \psi_{3,n} \right> &= \sqrt{\frac{n-1}{4n-2}} \left| 0,n-2
 \right> -
\frac{1}{\sqrt{2}} \left| 1,n-1 \right> \nonumber \\ &+
\sqrt{\frac{n}{4n-2}} \left| 2,n \right>,
\end{align}
\end{subequations}
with energies
\begin{subequations}
\begin{align}
E_{1,n} &= \hbar\omega(n-1) + \Lambda\sqrt{4n-2}, \\
E_{2,n} &= \hbar\omega(n-1), \\
E_{3,n} &= \hbar\omega(n-1) - \Lambda\sqrt{4n-2}.
\end{align}
\end{subequations}

\subsection{Arbitrary Spin $\sigma$}\label{arbspin}

A simple analytic solution exists for arbitrary $\sigma$ in the asymptotic
regime of large $n$. Consider the $(2\sigma+1)$-dimensional invariant block of
(\ref{hamil}) at $\alpha=0$ formed by the basis vectors
$|2\sigma-m,n-m\rangle$, $m=0,1,\ldots,2\sigma$. It is tridiagonal with
elements
\begin{align}
&\langle 2\sigma-m,n-m|H|2\sigma-m,n-m\rangle = \hbar\omega(n-\sigma), \nonumber \\
&\langle 2\sigma-m,n-m|H|2\sigma-m,n-m-1\rangle  \nonumber \\ &
\hspace*{4.5cm} = \Lambda\sqrt{2\sigma(n-m)}.\nonumber
\end{align}
For $n\gg\sigma$ we can write
\begin{equation}\label{hn1}
H = \hbar\omega(n-\sigma)\mathcal{E} + 2\Lambda\sqrt{n}S_x 
    + {\rm O}\left(\frac{\sigma}{\sqrt{n}}\right),
\end{equation}
where $\mathcal{E}$ is the $(2\sigma+1)$-dimensional unit matrix and $S_x$ is the irreducible
representation of the spin operator with the same dimensionality. The asymptotic
eigenvalues of this matrix are
\begin{equation}
E_{m,n} \simeq \hbar\omega(n-\sigma) 
        + 2\Lambda\sqrt{n}(\sigma-m).
\end{equation}
for $m=0,\dots,2\sigma$. The corresponding analysis carried out for
$\alpha=\pi/2$ yields the matrix
\begin{eqnarray}\label{hn2}
H = \hbar\omega(n-\sigma)\mathcal{E} + 2\hbar\omega S_z  
+ 2\Lambda\sqrt{n}S_x + {\rm O}\left(\frac{\sigma}{\sqrt{n}}\right)
\end{eqnarray}
with asymptotic energy eigenvalues (for $m=0,\dots,2\sigma$)
\begin{equation}
E_{m,n} \simeq \hbar\omega(n-\sigma) 
        + 2\hbar\omega\sqrt{\lambda_n}(\sigma-m).
\end{equation}

Note that in all cases pertaining to the integrable regimes $\alpha=0$ or
$\alpha=\pi/2$ the energy levels are naturally labelled by the two quantum
numbers $m,n$. The parity becomes $P=(-1)^{m+n}$. In the nonintegrable regime
$0<\alpha<\pi/2$, by contrast, the numerical analysis suggests the use of a
single (energy sorting) quantum number $k$ for all levels of given
parity. 

%
\section{Quantum Invariants}\label{sec3}
%

The quantum counterparts of the two analytic invariants (\ref{cl-inv}) are the
operators
\begin{equation}\label{IKq}
I = \hbar(a^\dagger a + S_z),\quad K = \hbar(a^\dagger a - S_z),
\end{equation}
which indeed commute with (\ref{hamil}) under exactly the same conditions as in
the classical limit. We have
\begin{subequations}
\begin{align}
\left[H,I \right] &= 2\Lambda\sin\alpha(S_-a-S_+a^\dagger), \\ 
\left[H,K \right] &= 2\Lambda\cos\alpha(S_+a-S_-a^\dagger).
\end{align}
\end{subequations}
However, quantum integrability cannot be inferred from quantum
invariants as simply as classical integrability can be inferred from 
integrals of the motion (analytic invariants). Commuting operators
can always be constructed irrespective of whether the model is (classically)
integrable or not \cite{Weig92, WM95}. The parity operator (\ref{pari}), for
example, which can be expressed as a function of either invariant $I$ or $K$,
\begin{equation}
P = e^{i\pi(I/\hbar-\sigma)} =  e^{i\pi(K/\hbar+\sigma)},
\end{equation}
commutes with $H$ for arbitrary $\alpha$.  More generally, any operator $A$
that is not already an invariant, $[H,A]\neq 0$, can be turned into an
invariant via time average. In the energy representation, the time average
strips $A$ of all its off-diagonal elements.  The resulting operator $I_A =
\langle A \rangle$ thus commutes with $H$ by construction \cite{Pere84,SM90}.

The fact is that in the classical limit neither the parity operator nor any of
the artificially constructed quantum invariants will turn into analytic
invariants (integrals of the motion) if the phase flow is chaotic. Such quantum
invariants either lose their meaning altogether or turn into nonanalytic
invariants \cite{SKM+88,SM90}.

The distinctive attributes of quantum invariants in the integrable and
nonintegrable regimes of a quantum system are subtle but not
ambiguous. Here we use
\begin{equation}\label{ia}
I_A = \langle A \rangle,\quad A = a^\dagger(S_- + S_+).
\end{equation}
For $\sigma=\frac{1}{2}$, its eigenvalues at $\alpha=0$ can be calculated
from the eigenvectors (\ref{psi1}) ,
\begin{equation}\label{ia1}
\langle A \rangle_{1,n} = \frac{1}{2}\sqrt{n},\quad
\langle A \rangle_{0,n} = -\frac{1}{2}\sqrt{n+1},
\end{equation} 
and its eigenvalues at $\alpha=\pi/2$ from the eigenvectors (\ref{psi2}): 
\begin{subequations}\label{ia2}
\begin{align}
\langle A \rangle_{1,n} &= \frac{(\lambda_n-1)(\sqrt{\lambda_n}-1)}
{2(\lambda_n - \sqrt{\lambda_n})}, \\
\langle A \rangle_{0,n} &= -\frac{(\lambda_{n+1}-1)(\sqrt{\lambda_{n+1}}+1)}
{2(\lambda_{n+1} + \sqrt{\lambda_{n+1}})}. 
\end{align}
\end{subequations} 
For $\sigma=1$ and $\alpha=0$, we obtain the following eigenvalues from
the eigenvectors (\ref{evpi2}):
\begin{equation}
\langle A\rangle_{1,n} = -\langle A\rangle_{3,n} = \sqrt{n-1/2},\quad
\langle A\rangle_{2,n} = 0,
\end{equation}
Asymptotically for large $n$, we can evaluate the eigenvalues for
arbitrary $\sigma$. The results for $\alpha=0$ read 
\begin{equation}
\langle A\rangle_{m,n} = \left(\sigma-m\right)\sqrt{n},
\end{equation}
and for $\alpha=\pi/2$ we have
\begin{equation}
\langle A\rangle_{m,n} = 
\frac{(\sigma-m)\sqrt{n}}{\sqrt{1+1/n\lambda}}.
\end{equation}

Numerical results of $\langle A\rangle_k$ for $0<\alpha<\pi/2$ of systems with
$\sigma=\frac{1}{2},1,\frac{3}{2}$ were reported previously \cite{CCM95}. The
patterns of points $(E_{m,n},\langle A\rangle_{m,n})$ for integrable cases were
found to be strikingly different from the pattern of points $(E_k,\langle
A\rangle_k)$ for nonintegrable cases. Here this difference will be used as a
demarcation tool for regimes of integrability and nonintegrability.

%
\section{Quantum Actions}\label{sec4}
%

One hallmark of integrability in a quantum system with two degrees of freedom
is that the Hamiltonian can be expressed as a function of two action operators
$J_1,J_2$, i.e. of two quantum invariants whose spectra consist of equidistant
levels.

\subsection{From $\Lambda=0$ to $\Lambda>0$}\label{sec4a}

In the absence of the spin-boson interaction $(\Lambda=0)$, the two action
operators are
\begin{equation}\label{qa}
J_1 = \hbar(\sigma-S_z),\quad J_2 = \hbar a^\dagger a
\end{equation}
with integer eigenvalues (in units of $\hbar$)
\begin{subequations}\label{aqn}
\begin{align}
J_1 &= m\hbar,\quad m=0,1,\ldots,2\sigma, \\
J_2 &= n\hbar,\quad n=0,1,\ldots,
\end{align}
\end{subequations}
as in (\ref{on-dia}). The Hamiltonian, $H_0 =
\hbar\omega_B a^\dagger a + \hbar\omega_S S_z$, and the two quantum invariants
(\ref{IKq}) are expressible as linear combinations of $J_1,J_2$.

Classically, the contribution of each degree of freedom to $H_0 = p^2/2M +
\frac{1}{2}M\omega_B^2x^2 + \omega_S S_z$ is transformed into a function of one
action coordinate by a separate canonical transformation: $(S_z,\varphi) \to
(J_1,\theta_1)$ with $S_z = s-J_1$, $\varphi = -\theta_1$; and $(p,x) \to
(J_2,\theta_2)$ with $p = \sqrt{2J_2M\omega_B}\cos\theta_2$, $x =
\sqrt{2J_2/M\omega_B}\sin\theta_2$. The transformed Hamiltonian and the two
classical invariants (\ref{cl-inv}) are linear functions of $J_1,J_2$ just
as in quantum mechanics. The exact quantum spectra of $H_0,I,K$ can then be
recovered exactly via semiclassical quantization, i.e.  by substituting the
actions quantized according to (\ref{aqn}) into the classical Hamiltonian.


Classically, the interaction renders the equations of motion, Eqs.~(\ref{jc}),
nonlinear. However, the effects of anharmonicity in the time evolution depend
sensitively on whether integrability is sustained or destroyed by the
interaction. Integrability for $\alpha=0, \pi/2$ dictates that the phase flow
is exclusively toroidal. For $0<\alpha<\pi/2$ chaotic phase flow is omnipresent
albeit constrained by surviving tori.

Quantum mechanically, the interaction distorts the eigenvalue spectrum and
modifies the selection rules of transition rates. Quantum properties that are
as sensitive to the integrability status as their classical counterparts do
exist and have previously been explored in the context of a different model
system \cite{SM98,SM00,SM01}.

These properties are directly related to the existence of action operators as
constituent elements of the Hamiltonian such as discussed in Sec.~\ref{sec4a}
for the noninteracting system. In the interacting cases, the existence of
action operators can again be demonstrated directly for $\alpha =0, \pi/2$, and
their nonexistence for $0<\alpha<\pi/2$ can be demonstrated indirectly.

\subsection{$\sigma=\frac{1}{2},~ \alpha=0$}\label{sec4c}

The unitary transformation which diagonalizes the Hamiltonian (\ref{hamil}) for
$\sigma=\frac{1}{2}$ and $\alpha=0$, expressed in terms of spin and boson
operators, reads
\begin{equation}\label{U}
U_{A} = P_0^A + \frac{1}{\sqrt{2}} \left(-2 {S}_z + 
\frac{1}{\sqrt{{a}^{\dag} {a}}} {a}^{\dag} {S}_{-} +
{a} {S}_{+} \frac{1}{\sqrt{{a}^{\dag} {a}}} Q_1^A\right),  
\end{equation}
where $P_0^A = |1,0\rangle\langle1,0|$, $Q_1^A = 1- |0,0\rangle\langle0,0| -
|1,0\rangle\langle1,0|$. The operators
\begin{subequations}\label{tzbb0} 
\begin{align}
{T}_z & =  U_AS_zU_A^{-1} = P_0^AS_z - \frac{1}{2}G_1^A \\
{b}^{\dag} {b} & =  U_Aa^{\dag} aU_A^{-1} =  
{a}^{\dag} {a} - S_zP_0^A + \frac{1}{2}G_1^A 
\end{align}
\end{subequations}
with
\begin{equation}
G_1^A = {a} {S}_{+} \frac{1}{\sqrt{{a}^{\dag} {a}}} Q_1^A +
\frac{1}{\sqrt{{a}^{\dag} {a}}} {a}^{\dag} \hat{S}_{-}  
\end{equation}
are diagonal in the energy representation:
\begin{subequations}\label{tzbb0ev}
\begin{align}
 T_z|\psi_{m,n}\rangle &= (\sigma-m)|\psi_{m,n}\rangle,\\
 b^{\dag} b|\psi_{m,n}\rangle &= n|\psi_{m,n}\rangle.
\end{align}
\end{subequations}
Hence the quantum actions with eigenvalues (\ref{aqn}) are
\begin{equation}\label{qa0}
J_1 = \hbar(\sigma-T_z),\quad  J_2 = \hbar b^{\dag} b.
\end{equation}
Applying $U_A$ to the Hamiltonian yields
\begin{align}\label{hjj0}
U_AHU_A^{-1} &= \hbar\omega(b^{\dag} b + T_z) \nonumber \\
&\hspace*{-7mm}+ \Lambda\left[\frac{1-2T_z}{2}\sqrt{b^{\dag} b} -
\frac{1+2T_z}{2}\sqrt{b^{\dag} b+1}\right],
\end{align} 
which, together with (\ref{qa0}), describes the functional relation between
$H$ and $J_{1},J_{2}$.

\subsection{$\sigma=\frac{1}{2},~ \alpha=\pi/2$}\label{sec4d}

The same method also produces the quantum actions for the integrable case
$s=\frac{1}{2}, \alpha=\pi/2$ of the spin-boson Hamiltonian (\ref{hamil}). Here
the block-diagonal unitary transformation, $U_B$ to be used can also
be expressed in terms of spin and boson operators but has a more complicated
structure than $U_{A}$. The operators
\begin{subequations}\label{tzbb1}
\begin{align}
{T}_z &= U_BS_zU_B^{-1} = \frac{1}{2}P_0^B +\left(G_B \right))Q_0^B , \\
{b}^{\dag} {b} &= U_Ba^{\dag} aU_B = \left({a}^{\dag} {a} - {S}_z + G_B \right)Q_0^B ,
\end{align} 
\end{subequations}
with $G_B=G_1^B+G_2^B+G_3^B$,
\begin{align*}
G_1^B &= \frac{1} {\sqrt{16 \left(1+\lambda {a}^{\dag} {a}\right)}} -
\frac{1}{\sqrt{16 \left(1+\lambda \left(
{a}^{\dag} {a} +1 \right) \right)}}, \\
G_2^B &= \frac{S_z} {\sqrt{4 \left(1+\lambda {a}^{\dag} {a} \right)}} +
\frac{S_z} {\sqrt{4 \left(1+\lambda \left( {a}^{\dag} {a} + 1 
\right) \right) }}, \\
G_3^B &= \frac{ 1+2 {S}_z} {4 \sqrt{64 \lambda + {a}^{\dag} {a}}}
{a}^{\dag} {S}_{+} + {a} {S}_{-} \frac{1+ 2{S}_z}{4 \sqrt{ 64 \lambda + 
{a}^{\dag} {a}}},
\end{align*}
and $P_0^B = |0,0\rangle\langle0,0|$, $Q_0^B = 1 - |0,0\rangle\langle0,0|$,
again satisfy (\ref{tzbb0ev}) and are related to quantum actions via
(\ref{qa0}). The functional dependence of the transformed Hamiltonian on the
actions is different from (\ref{hjj0}):
\begin{align}\label{hamb}
U_BHU_B^{-1} &= \hbar\omega(b^{\dag} b-T_z) + 
\frac{1+2T_z}{2}\sqrt{1+\lambda b^{\dag} b} \nonumber \\
&\hspace*{10mm}- \frac{1-2T_z}{2}\sqrt{1+\lambda(b^{\dag} b+1)}.
\end{align}

\subsection{$\sigma>\frac{1}{2}$}\label{sec4e}

The results of Secs.~\ref{sec4c} and \ref{sec4d} are generalizable to arbitrary
$\sigma$, albeit for the price of a higher and higher calculational effort.
The case $\sigma=1, \alpha=0$ can still be presented compactly. The unitary
transformation $U_C$ to be used in this case is now determined by the
eigenvectors (\ref{evpi2}) and yields
\begin{subequations}\label{tzbb2}
\begin{align}
T_z &= U_CS_zU_C^{-1} = P_0^C - G_1^CQ_2^C + \frac{1}{2}G_2^CP_1^C, \\
b^{\dag} b &= U_Ca^{\dag} aU_C^{-1} \nonumber \\
&= \left(a^{\dag} a + S_z +G_1^C \right)Q_1^C +
\frac{1}{2}G_2^CP_1^C,
\end{align}
\end{subequations}
where 
\begin{align*}
G_1^C &= \frac{S_z^2-S_z}{2\sqrt{4a^{\dag} a-2}}a^{\dag} S_- 
+ \frac{1-S_z^2}{\sqrt{4a^{\dag} a+2}}aS_+ \\
&+ \frac{1-S_z^2}{\sqrt{4a^{\dag} a+2}}a^{\dag} S_- 
+ \frac{S_z+S_z^2}{\sqrt{4a^{\dag} a+6}}a^{\dag} S_+, \\
G_2^C &= 1 + \frac{1-S_z^2}{\sqrt{2}}aS_+ 
+ \frac{S_z^2-S_z}{2\sqrt{2}}a^{\dag} S_-, 
\end{align*}
and $P_0^C = |0,0\rangle\langle0,0|$, $P_1^C = |1,0\rangle\langle1,0| + 
|0,1\rangle\langle0,1|$, $Q_2^C = 1 - P_0^C - P_1^C$.
The transformed Hamiltonian becomes
\begin{align}\label{hamc}
U_CHU_C^{-1} &= \hbar\omega\left(b^{\dag} b + T_z \right) 
+ \frac{\Lambda}{\sqrt{2}}
\left[\left(3T_z^2 - S_z -2 \right)P_1^C  \right. \nonumber \\
&\hspace*{-15mm+ \left. \left( (T_z^2-T_z)\sqrt{2b^{\dag} b-1} 
}-   (T_z^2+T_z)\sqrt{2b^{\dag} b+3}\right)Q_2^C\right].
\end{align}

$U_A$ and $U_C$ are special cases for $\sigma=\frac{1}{2}$ and $\sigma=1$,
respectively, of a unitary transformation $U_1(\sigma,\Lambda)$ that
diagonalizes (\ref{hamil}) at $\alpha=0$ for arbitrary values of $\sigma$. This
transformation turns out to be $\Lambda$-independent for the two cases we have
worked out. It may well be $\Lambda$-independent for arbitrary $\sigma$.
Likewise, $U_B$ is the special case for $\sigma=\frac{1}{2}$ of a unitary
transformation $U_2(\sigma,\Lambda)$ that diagonalizes (\ref{hamil}) at
$\alpha=\pi/2$ for arbitrary $\sigma$. That transformation is manifestly
$\Lambda$-dependent.

The end-product of these unitary transformations are two functions
$\bar{H}_Q^{(1)}(T_z,b^{\dag} b;\Lambda) = H_Q^{(1)}(J_1,J_2;\Lambda)$ and
$\bar{H}_Q^{(2)}(T_z,b^{\dag} b;\Lambda) = H_Q^{(2)}(J_1,J_2;\Lambda)$, which
express the functional dependence of the Hamiltonian on action operators in the
two integrable regimes $\alpha=0$ and $\alpha=\pi/2$, respectively.  The
leading term of an asymptotic expansion at high boson occupancy and
unrestricted spin state of these functions can be inferred from (\ref{hn1}) and
(\ref{hn2}):
\begin{equation}\label{hqbar}
  \bar{H}_Q = \hbar\omega\left(b^{\dag} b \pm T_z \right) + 
\Lambda \sqrt{b^{\dag} b} + {\rm O}(1),
\end{equation}
where the operators $T_z, b^{\dag}b$ again satisfy (\ref{tzbb0ev}) and the
upper (lower) sign pertains to $\alpha=0$ $(\alpha=\pi/2)$.

We expect a semiclassical regime to exist at large spin
and/or boson quantum numbers where the functions $H_Q^{(1)}(J_1,J_2;\Lambda)$
and $H_Q^{(2)}(J_1,J_2;\Lambda)$ connect with functions
$H_C^{(1)}(J_1,J_2;\Lambda)$ and $H_C^{(2)}(J_1,J_2;\Lambda)$ of classical
actions. However, the identification of the semiclassical regime
requires a complete solution of the classical equations of motion (\ref{jc}), a
task still outstanding.

The connections between the quantum and classical functional dependences of
Hamiltonian on actions was investigated in a previous study for an integrable
two-spin model and for the (integrable) circular billiard model \cite{SM01}.
There we found subtle quantum effects that restrict the range of the
semiclassical regime in unexpected ways. That may also be the case in the
spin-boson model. However, the point we wish to emphasize in this study is a
different one.

%
\section{Tracking Eigenstates}\label{sec5}
%

The goal is to demonstrate that the functions $H_Q^{(1)}(J_1,J_2;\Lambda)$ and
$H_Q^{(2)}(J_1,J_2;\Lambda)$ cannot be extended in any consistent way into the
region of nonintegrability in the $(\Lambda,\alpha)$-plane. The functions
$H_Q^{(1)}$ and $H_Q^{(2)}$ make it possible to label all eigenstates of
(\ref{hamil}) by the two action quantum numbers $m,n$ as defined in (\ref{aqn})
and to track them with no ambiguity through each one of the two integrable
regimes $\alpha=0$ and $\alpha=\pi/2$.  The non-extendability of the two
functions $H_Q^{(1)}$ and $H_Q^{(2)}$ into a function
$H_Q(J_1,J_2;\Lambda,\alpha)$ translates into the impossibility of consistently
assigning action quantum numbers $m,n$ to the eigenstates in the entire
parameter range $0\leq\Lambda<\infty$, $0\leq\alpha\leq\pi/2$.

One way of keeping track of eigenstates $|\nu\rangle$ of (\ref{hamil}) is to
determine how the eigenvalues of quantum invariants vary along some path in the
$(\Lambda,\alpha)$-plane. For the purpose of this demonstration, we focus on
the eigenvalues $\langle H\rangle_\nu=E_\nu$ of the Hamiltonian (\ref{hamil})
with spin quantum number $\sigma=\frac{1}{2}$ and the eigenvalues $\langle
A\rangle_\nu$ of the quantum invariant $I_A$ as defined in (\ref{ia}).

Within each of the two integrable regimes, both sets of eigenvalues have an
explicitly known (discrete) dependence on the action quantum numbers
$\nu=(m,n)$ and an explicitly known (continuous) dependence on the interaction
strength $\Lambda$. The functional relations are stated in Eqs. (\ref{emn1}),
(\ref{ia1}) for $\alpha=0$ and (\ref{emn2}), (\ref{ia2}) for $\alpha=\pi/2$.

\subsection{Level Crossings}\label{levcro}

In panels (a) and (b) of Fig.~\ref{fig3} we have plotted one quantum invariant
versus the other for all states with positive parity up to a certain energy. In
both panels we observe two vertically displaced rows of states. States in the
top and bottom rows have action quantum numbers $(1,n)$ and $(0,n)$,
respectively.

The observed arrangement of states is due to the fact that $\langle
A\rangle_{m,n}\sim\sqrt{n}$ but $E_{m,n}\sim n$ in leading order. Notice that
the spacings between successive energy levels in each row vary slowly, and at
different rates in the top and bottom rows.  To enhance the visibility of this
effect we have connected successive energy levels in each panel by dashed
lines. The spacings are somewhat larger in the top row compared to the bottom
row, causing instances in both panels where two consecutive states of the
bottom row fit into the space between two states of the top row. These
instances where the alternating (top/bottom) sequence is broken mark locations
where energy levels from opposite rows can fall arbitrarily close to each
other.

When we increase the interaction strength $\Lambda$ gradually, the states in
the top row of Fig.~\ref{fig3}(a) move toward the right and the states in the
bottom row toward the left. The same observation can be made in
Fig.~\ref{fig3}(b). Here the shift also contains a small vertical component.
We have singled out one pair of nearly degenerate states in Fig.~\ref{fig3}(a)
and another pair in Fig.~\ref{fig3}(b). Each pair is marked by full circles. In
panels (a) and (b) of Fig.~\ref{fig4} we have plotted the traces of these states 
in the plane of invariants as the interaction strength is increased by a certain 
amount.

The gradual change of $\Lambda$ causes a cascade of level crossings
between states from opposite rows. For the two pairs of tagged states, the
crossings occur at the point marked by an asterisk on their traces. States from
opposite rows undergo level crossings even though they have the same parity.
What matters are the functional relations $H_Q^{(1)}$ and $H_Q^{(2)}$
established previously.  They remove any possible cause for level collisions
(avoided crossings) between states from opposite rows as they move
(energetically) in opposite directions when $\Lambda$ is increased.

\subsection{Level Collisons}\label{levcol}
\begin{figure}[t!]
\centerline{\includegraphics[width=6.0cm,angle=-90]{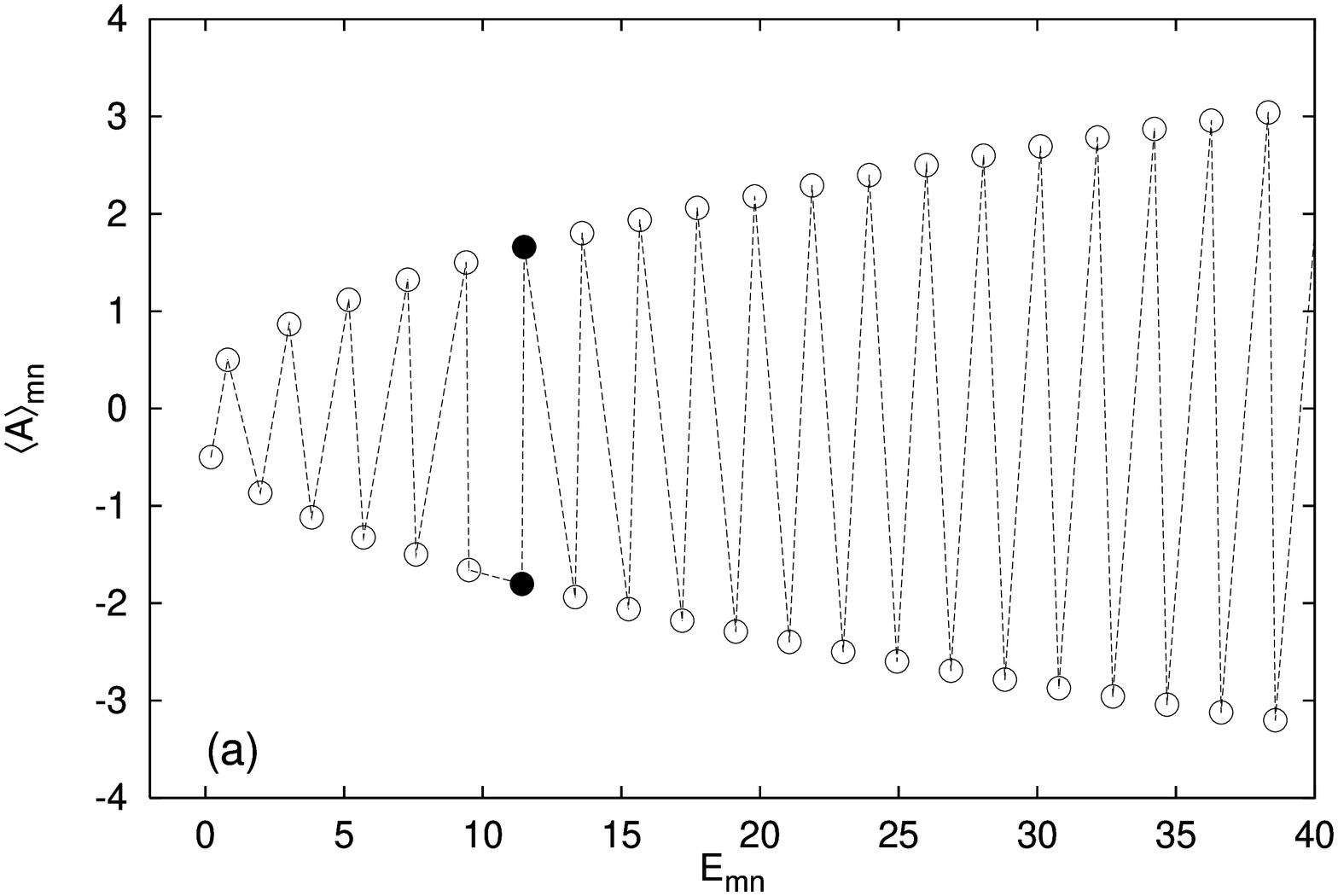}}
\centerline{\includegraphics[width=6.0cm,angle=-90]{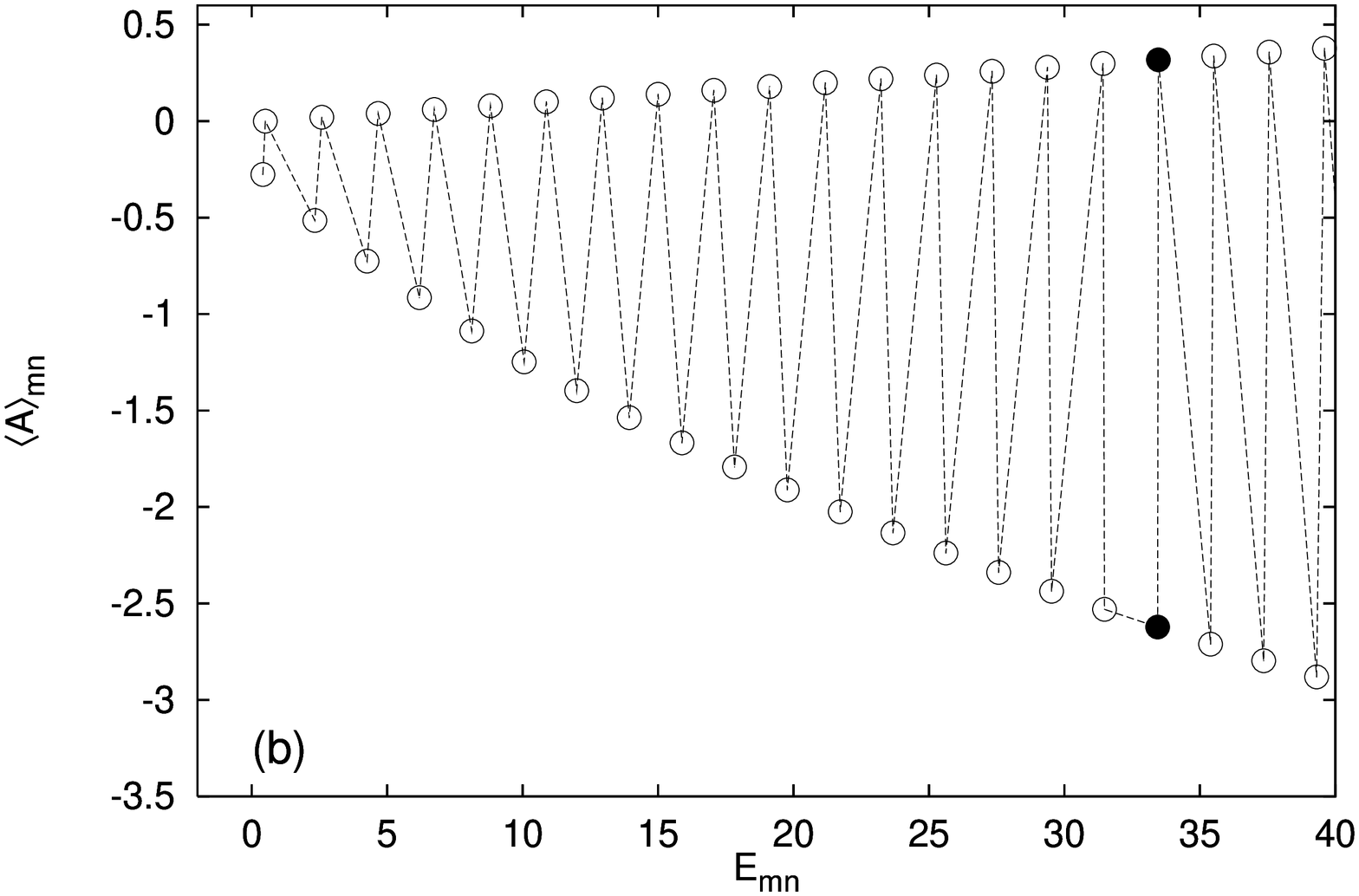}}
\centerline{\includegraphics[width=6.0cm,angle=-90]{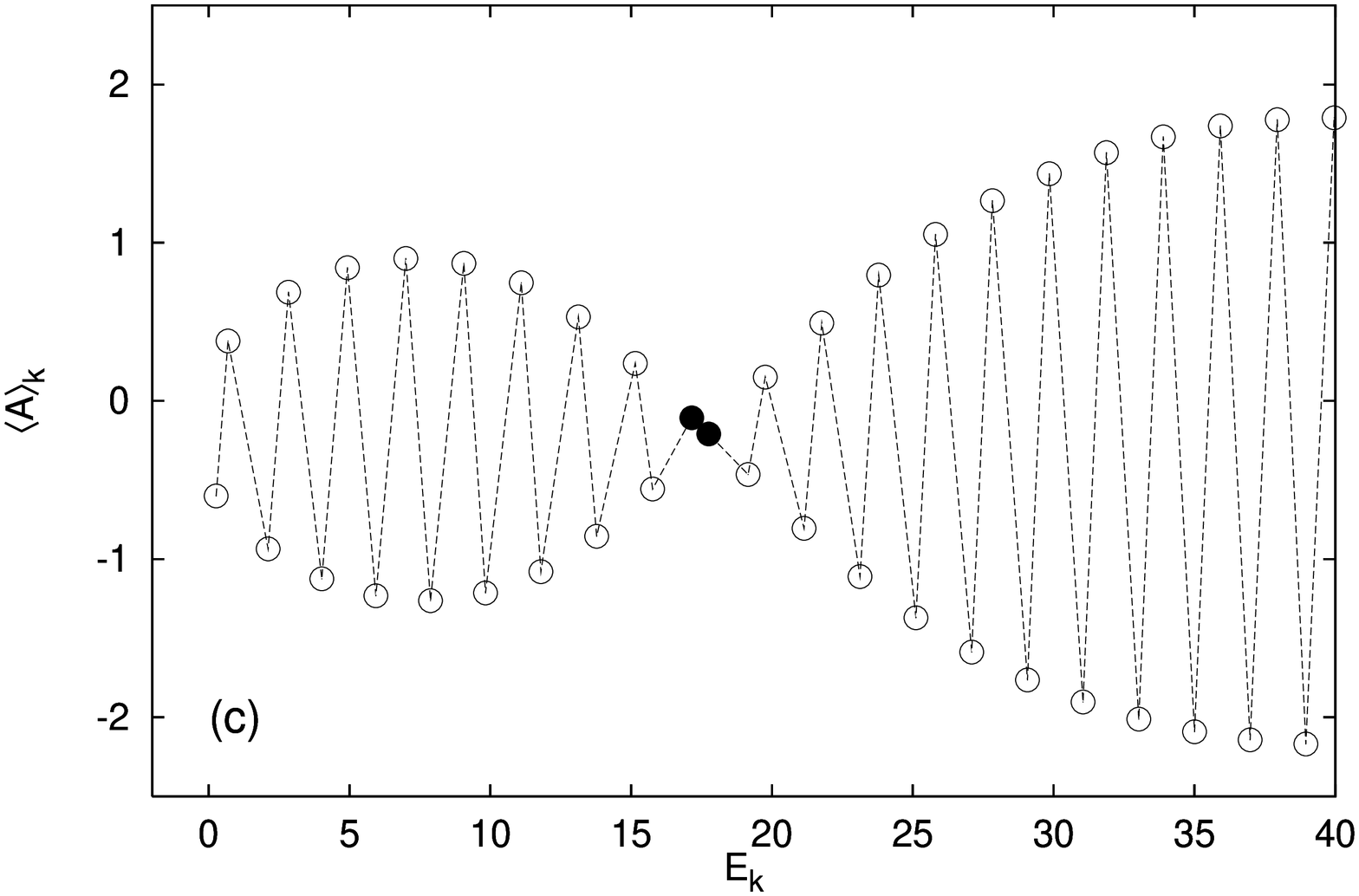}}
\caption{Quantum invariant $\langle A\rangle_\nu = \langle\nu|a^{\dag}(S_-
  +S_+)|\nu\rangle$ versus quantum invariant $E_\nu = \langle\nu|H|\nu\rangle$
  over some energy range for the eigenstates $|\nu\rangle$ with parity $P=+1$
  of the spin-boson model (\ref{hamil}) with $\sigma=\frac{1}{2}$,
  $\hbar\omega=1$, $\lambda\doteq(\Lambda/\hbar\omega)^2=0.09$, and (a)
  $\alpha=0$, (b) $\alpha=\pi/2$, (c) $\alpha=\pi/4$. In the integrable regimes
  we use the action quantum numbers $\nu=(m,n)$ and in the nonintegrable regime
  we use the energy sorting quantum number $\nu=k$. One pair of states in each
  panel (full circles) is tagged for further use in Fig.~\ref{fig4}. }
\label{fig3}
\end{figure}

A very different scenario unfurls when we plot the two quantum invariants for a
nonintegrable case. What happens when we change the integrability parameter
from $\alpha=0$ [Fig.~\ref{fig3}(a)] or from $\alpha=\pi/2$
[Fig.~\ref{fig3}(b)] to $\alpha=\pi/4$ is illustrated in
Fig.~\ref{fig3}(c).  Here the states that used to live in different worlds (top
row with action quantum number $m=1$ and bottom row with $m=0$) now suddenly
get into each other's way. Since they are prohibited from undergoing any level
crossings, it is now appropriate to label them by the energy sorting quantum
number $k$.

In those parts of the spectrum where the energy level spacings are large, the
loss of integrability has no visible effect on the quantum invariants. That is
the case near the left and right border areas of Fig.~\ref{fig3}(c). Here the
two rows of states remain largely intact.  However, near the center of the
panel, where small energy level spacings occur, the eigenvectors of nearly
degenerate levels affect each other strongly. The most conspicuous effect is a
strong vertical displacement of the two states from the row positions toward
each other. Less conspicuous in Fig.~\ref{fig3}(c) but of even greater
importance is the small horizontal displacement of the two nearly degenerate
states away from each other. The effect of nonintegrability is that energy
levels exert a short-distance repulsion on each other. At the same time,
expectation values in general and the quantum invariant $\langle A\rangle_{k}$
in particular tend to become less differentiated than they were in the
integrable case.

When we again increase the interaction strength $\Lambda$, now at fixed
$\alpha=\pi/4$ in the nonintegrable regime, we find that no levels with equal
parity ever undergo a crossing. As in the integrable cases, the states with
$\langle A\rangle_k>0$ have a tendency to move toward the right and the states
with $\langle A\rangle_k<0$ toward the left.

\begin{figure}[t!]
\centerline{\includegraphics[width=6.0cm,angle=-90]{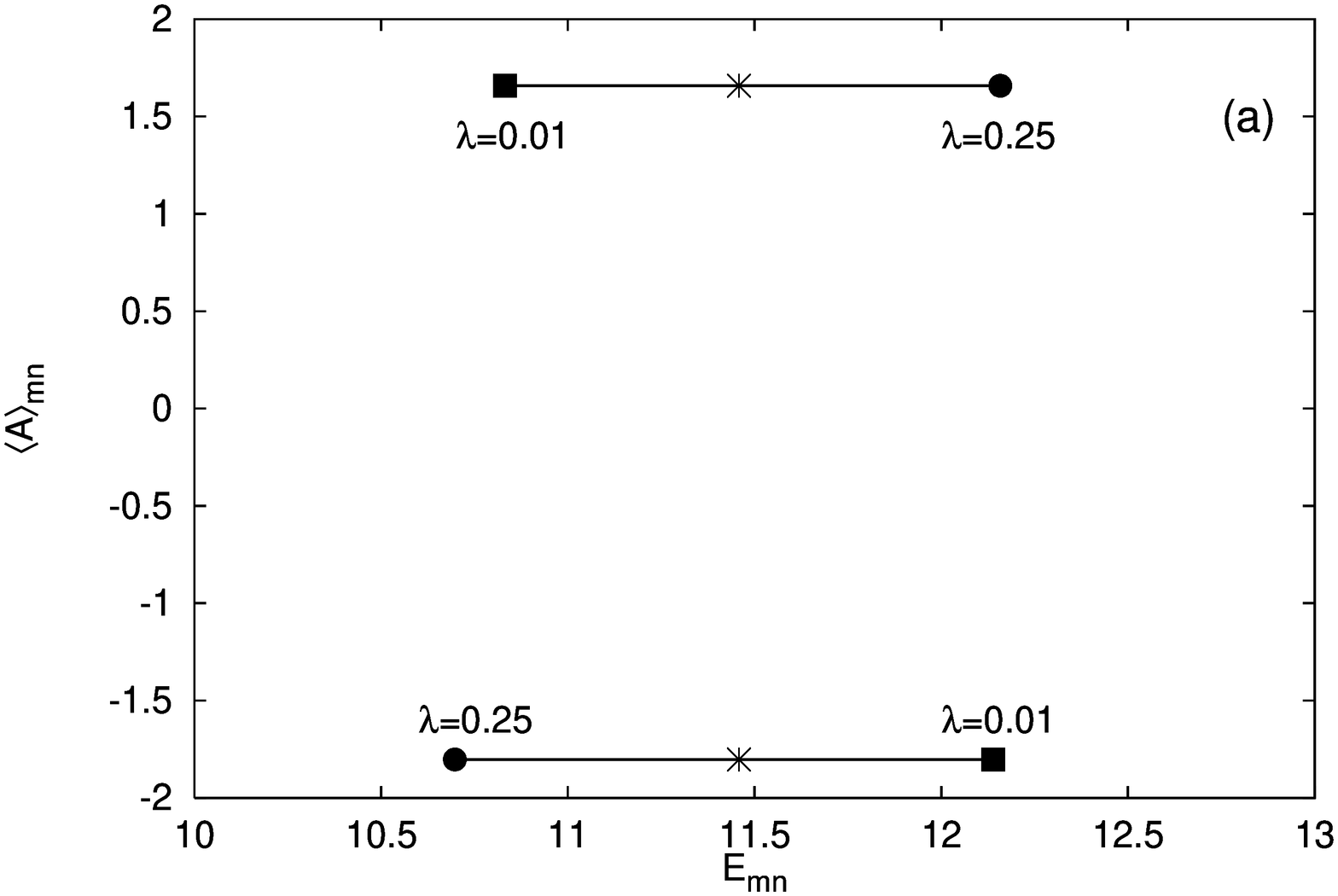}}
\centerline{\includegraphics[width=6.0cm,angle=-90]{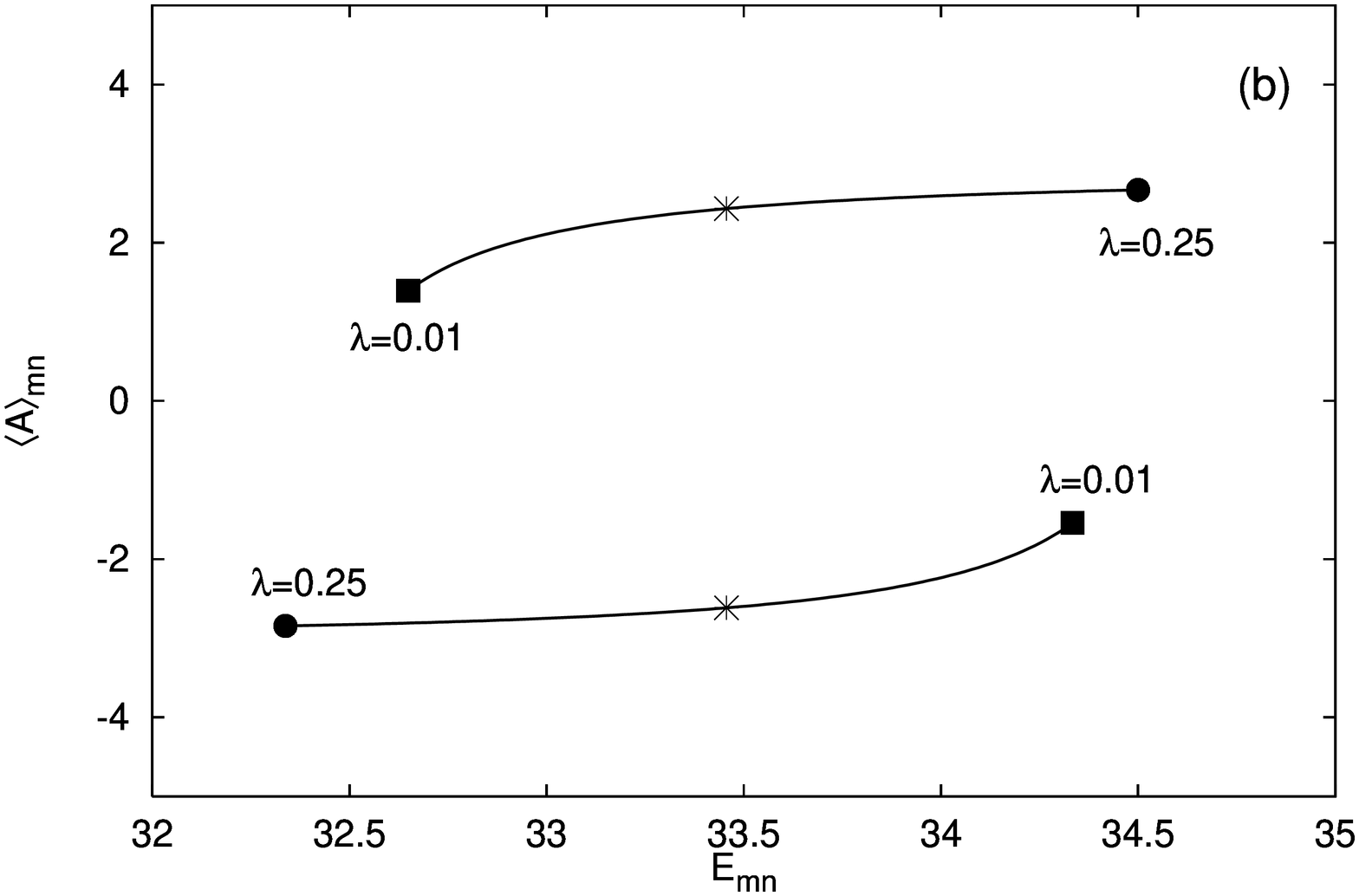}}
\centerline{\includegraphics[width=6.0cm,angle=-90]{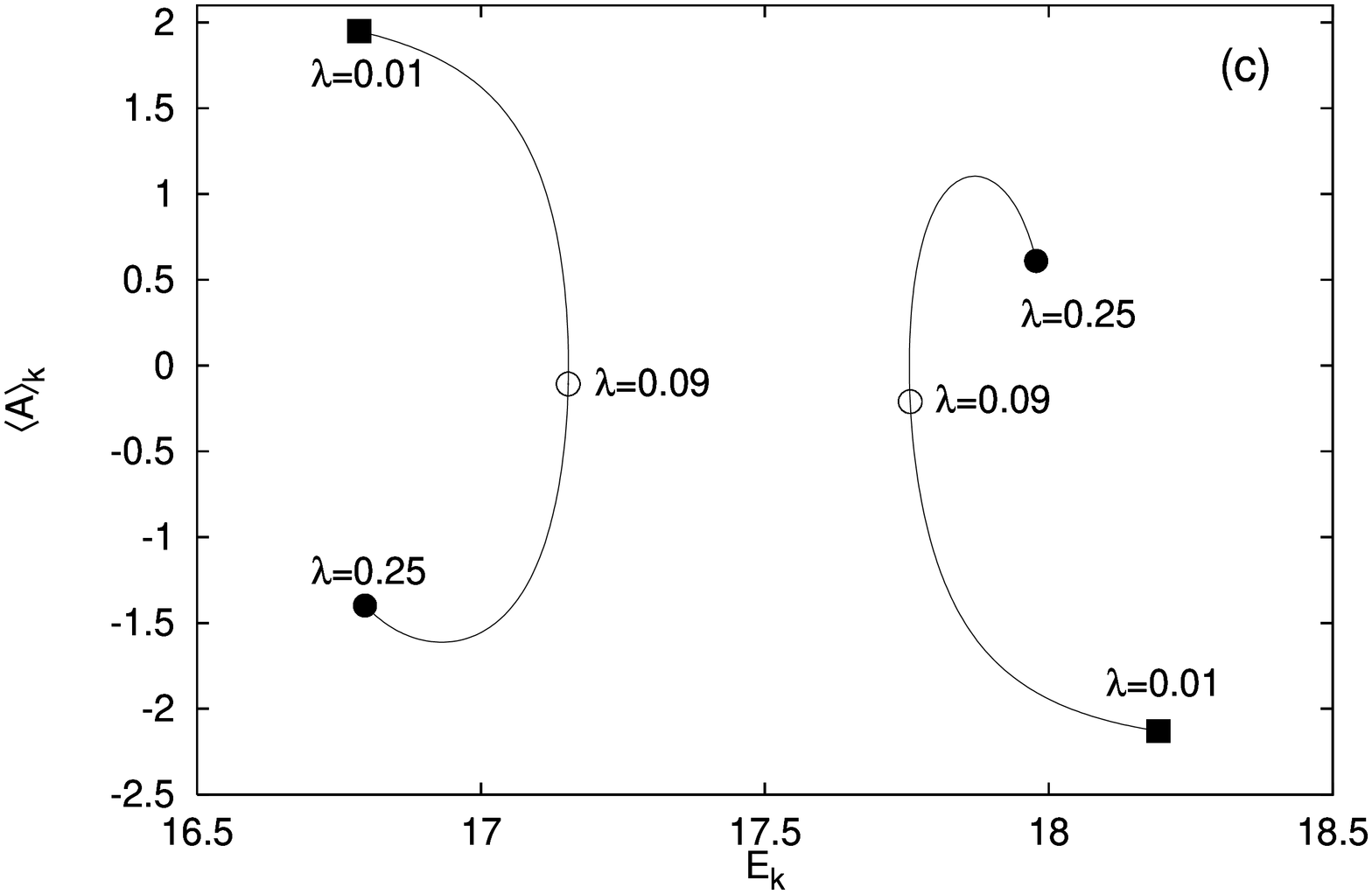}}
\caption{Trace of one pair of eigenstates $|\nu\rangle$ with parity $P=+1$
  (identified by full circles in Fig.~\ref{fig3}) in the plane of quantum
  invariants $(E_\nu,\langle A\rangle_\nu$ as the interaction parameter
  $\lambda$ is increased a specified amount at constant value (a) $\alpha=0$,
  (b) $\alpha=\pi/2$, (c) $\alpha=\pi/4$ of the integrability parameter. In the
  integrable regimes we use $\nu=(m,n)$ and in the nonintegrable regime
  $\nu=k$. }
\label{fig4}
\end{figure}

Inevitably, these trends put states on opposite sides of $\langle A\rangle_k=0$
on a collision course. When two such states approach one another, the state
starting out with $\langle A\rangle_k>0$ swings down as it moves to the right
and the state with $\langle A\rangle_k<0$ swings up as it moves to the left.
The two states reach their closest energetic approach when their vertical
positions are about the same. After that, the state coming from below continues
its upswing, but now it is moving to the right to join the right-moving upper
row of states. Meanwhile, the state coming from above continues its downswing
to join the left-moving lower row of states.  One such level collision, between
the tagged states in Fig.~\ref{fig3}(c), is shown in Fig.~\ref{fig4}(c).

\subsection{Quantum Numbers in Conflict}\label{qunuco}

Looking at the spectrum of the spin-boson model (\ref{hamil}) in the plane of
invariants $(E_\nu,\langle A\rangle_\nu)$ as the interaction strength $\Lambda$
increases gradually, reveals strikingly different patterns of coordinated
motion of all states with given parity, depending on whether the parameter
$\alpha$ is set to an integrable regime $(\alpha=0,\pi/2)$ or fixed within the
nonintegrable regime $(0<\alpha<\pi/2)$.

For $\alpha=0$ or $\alpha=\pi/2$ [panels (a) and (b), respectively, of
Figs.~\ref{fig3} and \ref{fig4}], the two rows of states march past each other
in an orderly fashion, undergoing a sequence of level crossings in complete
oblivion of each other's presence.
For $\alpha=\pi/4$ [panel (c) of Figs.~\ref{fig3} and \ref{fig4}], on the other hand,
all states are part of a coordinated clockwise looping motion. While every
individual state maintains the same position in the level sequence, the
wave-shaped top row of states has the appearance of moving steadily to the right 
and the bottom row to the left. The path of an individual state in the plane of
invariants is not unlike that of an H$_2$O molecule in a traveling surface water 
wave.

This qualitative change in pattern caused by different settings of the parameter 
$\alpha$ requires the assignment of mutually exclusive sets of quantum numbers to the 
same set of eigenstates in different parameter regimes. The action quantum
numbers $m,n$ are the trademark of quantum integrability. Their very existence
accommodates level crossings between states of equal parity. The level sorting
quantum number $k$, on the other hand, is applicable when level crossings
between states of equal parity are prohibited. It is the trademark of quantum
nonintegrability. 

This conclusion brings us full circle to the thought experiment on invariant
tori described in Sec.~\ref{sec1}. If we track an eigenstate along a closed
path in $(\Lambda,\alpha)$-plane, specifically a path that lies partly inside the
integrable regime and partly outside, its individuality cannot be maintained
through a unique and consistent assignment of quantum numbers. On a path that
first leads a certain stretch through the integrable regime and then returns
through the nonintegrable regime, the tagged eigenstate may undergo several
crossings on the first leg of this path and will then, on the second leg, 
be unable to cross back to its initial position in the level sequence.  
Barring a minor caveat (see Appendix~\ref{sec:app}) this conflict in the
assignment of quantum numbers to eigenstates is a dependable detecting device
for the demarcation of regimes of integrability and nonintegrability in quantum
systems with few degrees of freedom.

\begin{appendix}

%
\section{Point of higher symmetry}\label{sec:app}
%

Conflicts in the assignment of quantum numbers to eigenstates may arise for
reasons unrelated to nonintegrability. In a study of a two-spin system
\cite{SM00} two such causes were identified: (i) the presence of points of
higher symmetry inside the integrable regime; (ii) a multiple connectedness of
the integrable regime in the parameter space. Both causes are readily
identified as extraneous. In the context of the spin-boson model (\ref{hamil})
only the first cause comes into play.

In the following we describe one scenario where two eigenstates swap positions
in the level spectrum when tracked along a closed path in parameter space, a
path that does not leave the integrable regime. For this purpose we consider
(\ref{hamil}) with $\sigma=\frac{1}{2}$ in the extended parameter space
$(\Lambda,\omega_{S}, \omega_{B})$ at $\alpha=0$.  
The energy eigenvalues
\begin{equation}\label{e_plus_minus} 
E_{\pm} = \left( n+\frac 12  \right) \hbar \omega_B \pm \frac 12  
\sqrt{4 \Lambda^2 (n+1) + (\hbar \omega_S - \hbar \omega_B)^2}, 
\end{equation} 
and the eigenvectors
\begin{subequations}\label{ket_plus_minus}
\begin{align} 
|+\rangle &= \cos \phi |0,n\rangle + \sin \phi |1,n\rangle, \\
|-\rangle &= -\sin \phi |0,n\rangle + \cos \phi |1,n\rangle,
\end{align}
\end{subequations} 
depend on the angular variable  
\begin{equation}\label{tan_phi} 
\phi = \arctan\;\frac{E_+ - n \hbar \omega_B - \frac 12 \hbar 
\omega_S}{\Lambda \sqrt{n+1}}. 
\end{equation} 
The point of higher symmetry is at $\Lambda=0, \omega_B=\omega_S$. Here the
energy eigenvalues become doubly degenerate (for $\omega_{B}>0$). 
We consider the quantum invariant 
\begin{equation} \label{def_inv} 
\langle S_z \rangle_{\pm} = \pm \frac 12 \cos 2\phi 
\end{equation} 
defined by expectation values in the eigenstates (\ref{ket_plus_minus}). 
The loop in parameter space is parametrized as follows:
\begin{subequations}\label{def_path} 
\begin{align}
 \hbar \omega_S &= \hbar \omega_B (1 + \sin \beta), \\
\Lambda &= \hbar \omega_B (1 - \cos \beta), 
\end{align}
\end{subequations}
where $0 \leq \beta \leq 2 \pi$. It cuts through the point of higher symmetry
at $\beta=0$. The crucial point is that one complete loop along this path
advances the angle (\ref{tan_phi}) by $\Delta\phi=\pi/2$, which interchanges
the two states (\ref{ket_plus_minus}) and does not bring both invariants
(\ref{e_plus_minus}) and (\ref{def_inv}) back to the same position. It takes
two loops to return the states $|\pm\rangle$ to their original identity and the
points $(E_{\pm},\langle S_z \rangle_{\pm})$ to their original position.

\end{appendix}

%
%



\begin{thebibliography}{100}
\bibitem{note1} 
The fundamental Poisson brackets for the physical variables are $\{p,x\}=1,
  \{S_\alpha,S_\beta\} = \sum_\gamma \epsilon_{\alpha\beta\gamma}S_\gamma$.
\end{thebibliography}
\end{document}